\documentclass[conference]{IEEEtran}
\IEEEoverridecommandlockouts
\usepackage{cite}
\usepackage{amsmath,amssymb,amsfonts}
\usepackage{algorithmic}
\usepackage{graphicx}
\usepackage{textcomp}
\usepackage{xcolor}
\def\BibTeX{{\rm B\kern-.05em{\sc i\kern-.025em b}\kern-.08em
    T\kern-.1667em\lower.7ex\hbox{E}\kern-.125emX}}

\usepackage{xspace}
\usepackage{mathtools}
\usepackage{booktabs}
\usepackage[ruled, linesnumbered]{algorithm2e}
\usepackage{stfloats}
\usepackage{subcaption}
\usepackage{bm}

\newcommand{\norm}[1]{\left\lVert#1\right\rVert}
\DeclareMathOperator*{\argmin}{arg\,min}
\DeclareMathOperator*{\argmax}{arg\,max}
\DeclarePairedDelimiter{\ceil}{\lceil}{\rceil}

\newcommand{\knn}[0]{kNN\xspace}
\newcommand{\dknn}[0]{DkNN\xspace}
\newcommand{\papernot}[0]{Papernot \& McDaniel\xspace}

\begin{document}

\title{On the Robustness of Deep K-Nearest Neighbors}

\author{
\IEEEauthorblockN{Chawin Sitawarin, David Wagner}
\IEEEauthorblockA{EECS Department, UC Berkeley}
\{chawins, daw\}@berkeley.edu
}

\maketitle

\begin{abstract}
Despite a large amount of attention on adversarial examples, very few works have demonstrated an effective defense against this threat. We examine Deep k-Nearest Neighbor (\dknn), a proposed defense that combines k-Nearest Neighbor (\knn) and deep learning to improve the model's robustness to adversarial examples. It is challenging to evaluate the robustness of this scheme due to a lack of efficient algorithm for attacking \knn classifiers with large $\bm{k}$ and high-dimensional data. We propose a heuristic attack that allows us to use gradient descent to find adversarial examples for \knn classifiers, and then apply it to attack the \dknn defense as well. Results suggest that our attack is moderately stronger than any naive attack on \knn and significantly outperforms other attacks on \dknn.
\end{abstract}



\section{Introduction}

Deep learning has recently attained immense popularity from various fields and communities due to its superhuman performance on complicated tasks such as image classification  \cite{krizhevsky12imagenet, simonyan14very}, playing complex games \cite{silver2017alphago, silver17zero, mnih13atari}, controlling driverless vehicles \cite{chen15deepdrive, bojarski16nvidia}, and medical imaging \cite{litjens17medical}. Nonetheless, many works have shown that neural networks and other machine learning classifiers are not robust in the face of adversaries (e.g. adversarial examples) \cite{biggio13, szegedy13, goodfellow14explaining, moosavi15deepfool, nguyen15} as well as more common cases of distribution shifts \cite{engstrom17, hendrycks18}. 

This phenomenon raises a call for more robust and more interpretable neural network models. Many defenses against adversarial examples have been proposed; however, most have been broken by adaptive adversaries \cite{carlini16distill, carlini17bypass, athalye18}. Only a few defenses provide a significant improvement in robustness on toy datasets like MNIST and CIFAR-10 \cite{madry17, xu17squeeze}. One plausible approach to simultaneously combat adversaries and make neural networks more trustworthy is to build interpretable models \cite{kim17, zhang17, papernot18dknn} or to provide an \textit{explanation} supporting the model's output \cite{simonyan13saliency, ribeiro16lime, guo18lemna}. Deep k-Nearest Neighbors (\dknn), recently proposed by \papernot, showed promising results: their evaluation suggests it offers robustness against adversarial examples, interpretability, and other benefits \cite{papernot18dknn}.

Nonetheless, adversarial examples are surprisingly difficult to detect when that the adversary has full knowledge of the defense \cite{carlini17bypass}. Among the works that have been beaten, many attempts to distinguish adversarial inputs by statistically inspecting their representation (or activation) from hidden layers of neural networks \cite{hendrycks16, li16, feinman17, grosse17detect}. This fact raises some concerns for the robustness of \dknn, which uses \knn on the intermediate representations produced by the neural network.

\begin{figure}[t]
  \centering
  \includegraphics[width=0.48\textwidth]{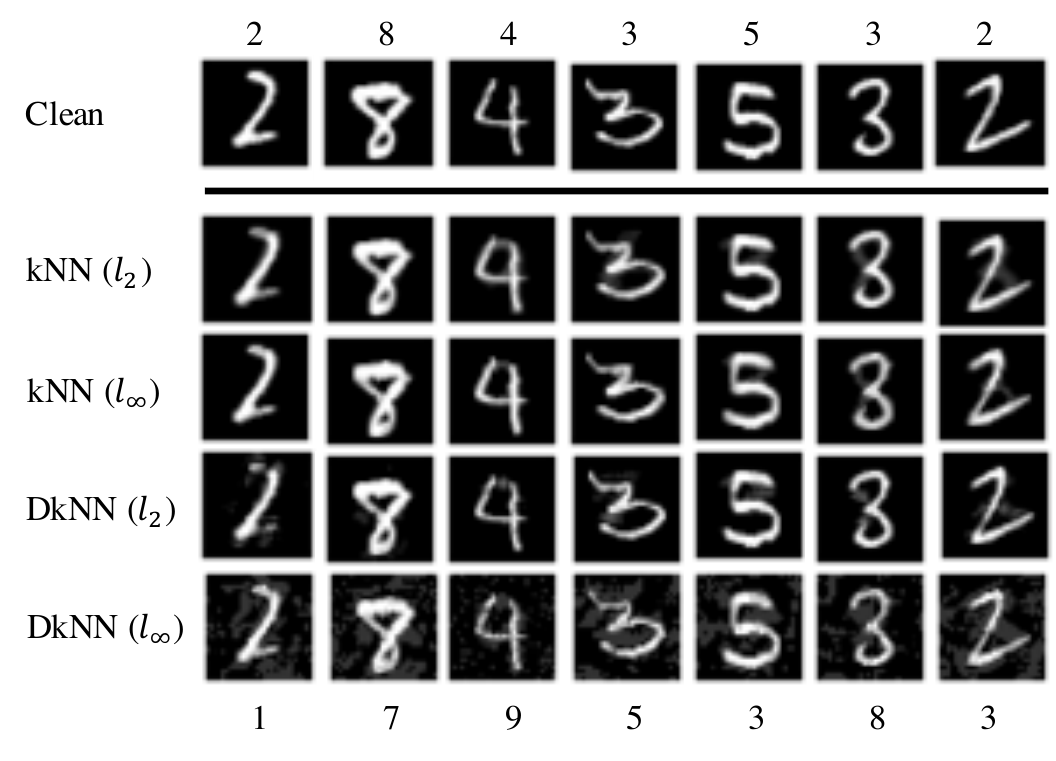}
  \vspace{-5pt}
  \caption{Adversarial examples generated from the gradient-based attack on \knn and \dknn with $\ell_2$- and $\ell_\infty$-norm constraints. The numbers on top and bottom are predictions of \dknn on the clean and the adversarial samples respectively. For a few adversarial examples, the perturbation might change the human label: some of the adversarial 4's have their top closed, so a human might consider them a 9, and one of the 3's looks close to an 8.}
  \label{fig:advex}
\end{figure}

In this paper, we examine the robustness of \dknn against adversarial examples.  We develop a new gradient-based attack on \knn and \dknn.  While gradient descent has found great success in attacking neural networks, it is challenging to apply to \knn, as \knn is not differentiable.  At a high level, our attack approximates the discrete nature of \knn with a soft threshold (e.g., a sigmoid), making the objective function differentiable. Then, we find a local optimum using gradient descent under an $\ell_p$-norm constraint. With this attack, we find that \dknn is vulnerable to adversarial examples with a small perturbation in both $\ell_2$ and $\ell_\infty$ norms. With $\ell_\infty$-norm of 0.2, our attack manages to reduce the accuracy of a \dknn on MNIST to only 17.44\%. Some of the adversarial examples generated with our attack are shown in Fig. \ref{fig:advex}.

The main contributions of this paper are as follows:
\begin{enumerate}
    \item We propose a gradient-based attack on \knn and \dknn. 
    \item We evaluate our attack on \knn and \dknn, compare it to other naive approaches as well as the adaptive attack proposed by \papernot, show that our attack performs better than prior attacks, and show that it can find adversarial examples for \knn and \dknn on MNIST.
    \item We show that the credibility scores from \dknn models are not effective for detecting our attacks without a significant drop in accuracy on clean images.
\end{enumerate}


\section{Background and Related Work}

\subsection{Adversarial Examples}

Adversarial examples are a type of an evasion attack against machine learning models at test time. While the robustness of machine learning classifiers in adversarial settings has been studied for a long time \cite{barreno06can, huang11advml}, the term ``adversarial examples'' was recently introduced as an attack on deep neural networks by adding very small perturbation to a legitimate sample \cite{szegedy13, goodfellow14explaining}. Previous works propose algorithms for finding such perturbation under a norm-ball threat model which can be generalized as solving the following optimization problem: 
\begin{align}
    x_{adv} = x + \delta^* \quad \text{where} ~ \delta^* = ~ &\argmax_{\delta} ~~ L(x + \delta) \\
    &\text{such that} ~ \norm{\delta}_p \leq d \nonumber
\end{align}
where $L$ is some loss function associated with the correct prediction of a clean sample $x$ by the target neural network. The constraint is used to keep the perturbation small or \textit{imperceptible} to humans. Our attack also uses the norm-ball constraint and an optimization problem of a similar form.

\subsection{Robustness of k-Nearest Neighbors}

The \knn classifier is a popular non-parametric classifier that predicts the label of an input by finding its $k$ nearest neighbors in some distance metric such as Euclidean or cosine distance and taking a majority vote from the labels of the neighbors. Wang et al. recently studied the robustness of \knn in an adversarial setting, providing a theoretic bound on the required value of $k$ such that robustness of \knn can approach that of the Bayes Optimal classifier \cite{wang18knn}. Since the required value of $k$ is too large in practice, they also propose a robust 1-NN by selectively removing some of the training samples. We did not experiment with this defense as it is limited to a 1-NN algorithm with two classes.

\subsection{Deep k-Nearest Neighbors}

\dknn, proposed by \papernot, is a scheme that can be applied to any deep learning model, offering interpretability and robustness through a nearest neighbor search in each of the deep representation layers. Using \textit{inductive conformal prediction}, the model computes, in addition to a prediction, \textit{confidence} and \textit{credibility} scores, which measure the model's assessment of how likely its prediction is to be correct. The goal is that adversarial examples will have low credibility and can thus be easily detected. The credibility is computed by counting the number of neighbors from classes other than the majority; this score is compared to scores seen when classifying samples from a held-out \textit{calibration set}. \papernot evaluate \dknn with an adaptive adversary which is found to be quite unsuccessful. We examine the robustness of \dknn with the stronger attack we propose.

We note that the \dknn proposed by \papernot uses cosine distance, which is equivalent to Euclidean distance given that all samples are normalized to have a unit norm. For the rest of the paper, we tend to omit the normalization for simplicity and less clutter in equations. The implementation and the evaluation, however, use cosine distance as instructed in the original paper.


\section{Threat Model}

We assume the white-box threat model for attacks on both \knn and \dknn. More precisely, the adversary is assumed to have access to the training set and all parameters of the \dknn neural network. Since a \knn classifier is non-parametric, the training set is, in some sense, equivalent to the weights of parametric models. We also assume that the adversary knows all hyperparameters, namely $k$, the distance metric used (Euclidean or cosine distance), and additionally the calibration set for \dknn. Though this knowledge is less crucial to the adversary, it allows the adversary to accurately evaluate his/her attack during the optimization resulting in a more effective attack.

For consistent comparisons with previous literature, the adversarial examples must be contained within a norm-ball ($\ell_2$ and $\ell_\infty$) centered at given test samples. We recognize that the $\ell_p$-norm constraint may not be representative of human perception nor applicable in many real-world cases.


\section{Attack on k-Nearest Neighbors} \label{sec: attack_knn}

\subsection{Notation}

We follow notation from \papernot as much as possible. Let $z$ denote a target sample or a clean sample that the adversary uses as a starting point to generate an adversarial example, and $y_z$  its ground-truth label. We denote the perturbed version of $z$ as $\hat{z}$. The training set for both \knn and \dknn is $(X, Y)$ with $n$ samples of dimension $d$. The classifier's prediction for a sample $x$ is $\mathsf{knn}(x)$.

\subsection{Mean Attack}

We first introduce a simple, intuitive attack to serve as a baseline.
Let $z$ be a clean sample, $y_z$ its ground-truth class, and $y_{adv} \ne y_z$ be a target class. The attack, which we call the mean attack, works by moving $z$ in the direction towards the mean of all samples in the training set with class $y_{adv}$. Concretely, we first search for the class $y_{adv} \ne y_z$ such that the mean of training samples with that class is closest to $z$ in Euclidean distance. Let $m$ denote the corresponding mean. We then use binary search to find the smallest $c>0$ such that $(1-c)z + cm$ is misclassified by the \knn. 



This attack is very simple to carry out and applicable to any classifier. While it is a natural choice for attacking a \knn with Euclidean distance, the attack may perform less well for cosine distance or other distance measures. As our experiments show, the mean attack also produces perturbations that make the resulting adversarial example look, to humans, more like samples from the target class, and thus makes the attack more noticeable. Nonetheless, this attack can be regarded as a simple baseline for measuring the robustness of nearest-neighbor classifiers.

\subsection{Naive Attack} 

\begin{figure}
  \centering
  \includegraphics[width=0.48\textwidth]{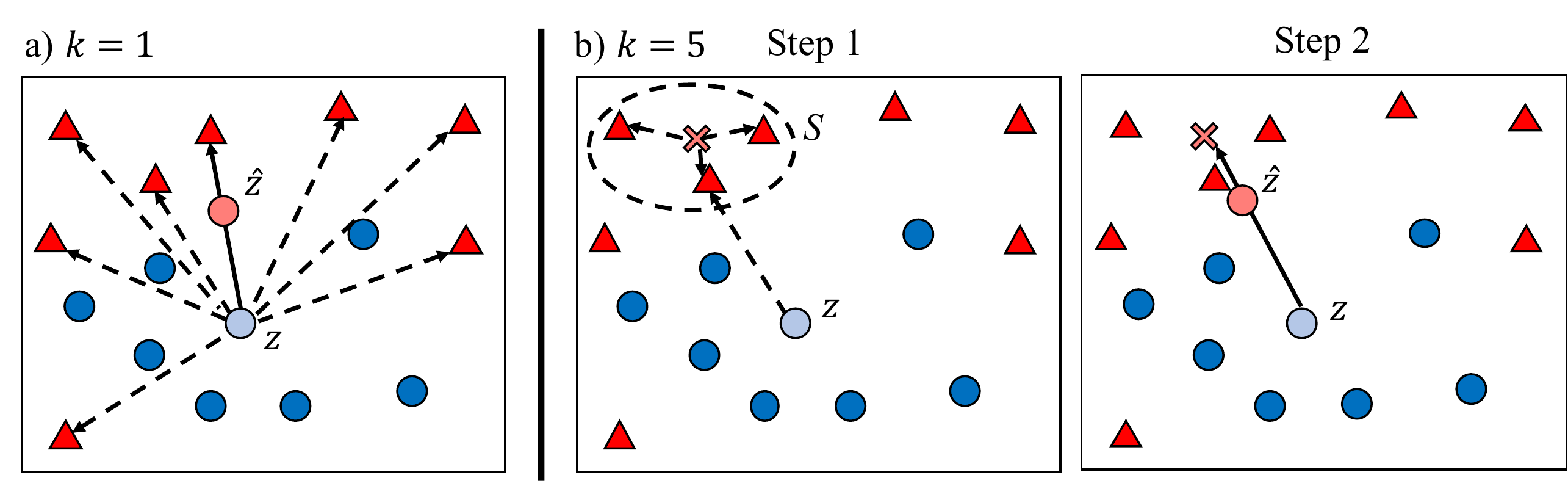}
  \caption{(a) naive attack for $k=1$: The target sample $z$ (light blue circle) is moved towards each of the samples from a different class (red triangles). The one that requires the smallest $\ell_2$-distance to change the prediction is the optimal adversarial example $\hat{z}$ (pink circle). (b) naive attack for $k > 1$: In the first step, a set $S$ of 3 samples from the different class closest to $z$ are located with a greedy algorithm. The second step involves moving $z$ towards a mean of the samples in $S$ and stops when the prediction changes.}
  \label{fig:naive}
\end{figure}

Next, we introduce a second baseline attack that improves slightly on the mean attack. When $k = 1$, a simple algorithm can find the optimal adversarial example in $\mathcal{O}(n)$ time. For each training sample $z'$ of a class other than $y_z$, the algorithm moves the target sample $z$ in a straight line towards $z'$ until $\mathsf{knn}(\hat{z}) \neq y_z$ (i.e., setting $\hat{z} = (1-c)z + cz'$, we find the smallest $c>0$ such that $\mathsf{knn}(\hat{z}) \neq y_z$). This produces $n$ candidate adversarial examples, and the algorithm outputs the one that is closest to $z$. Fig. \ref{fig:naive}(a) illustrates this algorithm. 

This strategy finds the optimal adversarial example when $k=1$, but when $k>1$, it is not clear how to find the optimal adversarial example efficiently. Repeating the previous strategy on all sets of $k$ training samples does not guarantee an optimal solution and is inefficient, as its complexity grows exponentially with $k$. Instead, we propose a computationally cheaper attack that greedily chooses only one set of samples to move towards, as summarized in Fig. \ref{fig:naive}(b). There are multiple possible heuristics to choose this set. One simple option would be to find the $\ceil{\frac{k}{2}}$ nearest neighbors of $z$ whose labels all match but are different from $y_z$. We instead use a slightly more complex variant: (1) find the nearest neighbor from any class other than $y_z$, say class $y_{adv}$, (2) add this sample to an empty set $S$, and (3) out of all samples with class $y_{adv}$, iteratively find the nearest sample to the mean of $S$ and add it to $S$. The final step is repeated until $|S| = \ceil{\frac{k}{2}}$. Finally, we move $z$ towards the mean of $S$ until the classifier's prediction differs from $y_z$.



\subsection{Gradient-Based Attack}

Here we introduce our main attack on \knn. On a high-level, it uses a heuristic initialization to choose a set of $m$ samples that are close to the target sample $z$. Then, a gradient-based optimization is used to move $z$ closer to the ones with the target class $y_{adv}$ and further from the ones with the original class $y_z$. 

We will discuss the choices for the heuristic initialization towards the end of this section. For now, the algorithm can be formulated as the following optimization problem. 
\begin{align} \label{eq: opt_knn_1}
    \hat{\delta} ~ = &\argmin_\delta ~ \sum_{i=1}^{m} ~w_i \cdot \norm{x_i - (z + \delta)}_2^2 \\
    &\text{such that} \norm{\delta}_p \leq \epsilon
    \text{ and } z + \delta \in [0, 1]^d \nonumber
\end{align}
where $\delta$ is the perturbation, $\hat{z} = z + \hat{\delta}$ is the adversarial example, $x_1,\dots,x_m$ are the $m$ training samples selected earlier, and $w_i = 1$ if the label of $x_i$ is $y_{adv}$, otherwise $w_i = -1$. The first constraint constrains the norm of the perturbation, and the second constraint ensures that the adversarial example lies in a valid input range, which here we assume to be $[0, 1]$ for pixel values. 

However, Eq. \ref{eq: opt_knn_1} may not achieve what we desire since it treats all $x_i$ equally and does not take into account that for \knn, only the $k$ nearest neighbors contribute to the prediction, while the other training samples are entirely irrelevant. Moreover, the distance to these $k$ neighbors does not matter as long as they are the $k$ closest. In other words, the distance to each of these $k$ neighbors is irrelevant so long as it is under a certain threshold $\eta$ (where $\eta$ is the distance to the $k$-th nearest neighbor). This means that a sample $x_i$ gets a vote if $\norm{x_i - \hat{z}}_2 \leq \eta$; otherwise, it gets zero vote. The optimization above does not take this into account.

We show how to adjust the optimization to model this aspect of \knn classifiers. The function that maps $\hat{z}$ to 0 or 1 according to whether $x_i$ gets a vote is not a continuous function and it has zero gradient where it is differentiable, so it poses challenges for gradient-based optimization. To circumvent this problem, we approximate the threshold with a sigmoid function, $\sigma(x) = \frac{1}{1 + e^{-\alpha x}}$ where $\alpha$ is a hyperparameter that controls ``steepness" (or an inverse of \textit{temperature}) of the sigmoid. As $\alpha \rightarrow \infty$, the sigmoid exactly represents the Heaviside step function, i.e., a hard threshold. This lets us adjust Eq. \ref{eq: opt_knn_1} to incorporate the considerations above, as follows:
\begin{align} \label{eq: opt_knn_2}
    \hat{\delta} ~ = &\argmin_\delta ~ \sum_{i=1}^{m} ~w_i \cdot \sigma \big( \norm{x_i - (z + \delta)}_2 - \eta \big)\\
    &\text{such that} \norm{\delta}_p \leq \epsilon
    \text{ and } z + \delta \in [0, 1]^d \nonumber
\end{align}
Ideally, $\eta$ should be recomputed at every optimization step, but this requires finding $k$ nearest neighbors at each step, which is computationally expensive. Instead, we fix the value of $\eta$ by taking the average distance, over all training samples, from each sample to its $k$-th nearest neighbor.

\textbf{Choosing the initial $\bm{m}$ samples.} There is no single correct way to initialize the set of $m$ samples. We empirically found that choosing all of them from the same class $y_{adv}$, and choosing the $m$ training samples of that class that are closest to $z$, works reasonably well. We choose $y_{adv}$ by computing the distance from $z$ to the mean of all samples of class $y$, for each $y$, and taking the class $y$ that minimizes this distance. Other heuristics might well perform better; we did not attempt to explore alternatives in depth, as this simple heuristic sufficed in our experiments. The choice of the attack parameter $m$ affects the attack success rate. A larger $m$ means we consider more training samples which make the \knn more likely to be fooled, but it is also more expensive to compute and may produce larger distortion. In principle, one could recompute the set of $m$ samples periodically as the optimization progresses, but for our experiments, we select them only once in the beginning.

For $p = \infty$, we use a change of variable as introduced by Carlini \& Wagner \cite{carlini17cw} to provide pixel-wise box constraints that simultaneously satisfy both of the optimization constraints in Eq. \ref{eq: opt_knn_2}. More precisely, the $i$-th pixel of the adversarial example is written as $\hat{z}_i = \frac{1}{2}(\tanh(v_i) + 1) \cdot (b_u - b_l) + b_l$ where $b_u$ and $b_l$ are the upper and the lower bound of that pixel respectively. $v$ becomes the variable that we optimize over, but for simplicity, we omit it from Eq. \ref{eq: opt_knn_2}. In the case of $p = 2$, this change of variables enforces the second constraint. The first constraint is relaxed and added to the objective function as a penalty term:
\begin{align} \label{eq: opt_knn_3}
\begin{split}
    \hat{\delta} ~ = \argmin_\delta ~ \sum_{i=1}^{m}& ~w_i \cdot \sigma \big(\norm{x_i - (z + \delta)}_2 - \eta\big) \\
    & + c \cdot \max\big\{0, ~\norm{\delta}_2^2 - \epsilon^2\big\} 
\end{split}\\
    \text{such that} \quad & z + \delta \in [0, 1]^d \nonumber
\end{align}
To find an appropriate value for $c$, we use a binary search for five steps. If the attack succeeds, $c$ is increased; otherwise, $c$ is decreased.


\section{Attack on Deep k-Nearest Neighbors}

\subsection{Notation}

Let $\mathsf{dknn}(x)$ denote \dknn's prediction for a sample $x$. The prediction of the $l$-layer neural network part of the \dknn is denoted as $f(x)$, and the output from the $\lambda$-th layer as $f_\lambda(x)$ where $\lambda \in \{1,2,...,l\}$. The calibration set $(X^c, Y^c)$ is used to calculate the empirical $p$-value as well as the credibility and confidence.

\subsection{Mean Attack}

The mean attack for \dknn is exactly the same as for \knn without any modification as the attack does not depend on the choice of classifiers.

\subsection{Baseline Attack}

We use the adaptive attack evaluated by \papernot as a baseline. Given a target sample $z$, we try to minimize the distance between its representation at the first layer and that of a \textit{guide sample} $x_g$, a sample from a different class whose representation is closest to $f_1(z)$. For the $\ell_\infty$-norm constraint, the attack can be written as:
\begin{align} 
    \hat{\delta} ~ = \argmin_\delta ~ &\norm{f_1(x_g) - f_1(z + \delta)}_2^2 \\ 
    \text{such that} \quad &\norm{\delta}_\infty \leq \epsilon
    \text{ and } z + \delta \in [0, 1]^d \nonumber
\end{align}
The optimization is solved with L-BFGS-B optimizer as suggested in Sabour et al. \cite{sabour15deeprep}. For completeness, we will also evaluate the attack with a $\ell_2$ constraint, using the same relaxation as Eq. \ref{eq: opt_knn_3}.

\subsection{Gradient-Based Attack} \label{ssec:dknn_grad}

The baseline attack relies on an assumption that if $f_1(\hat{z})$ is close to $f_1(x_g)$, then $f_\lambda(\hat{z})$ will also be close to $f_\lambda(x_g)$ for $2 \leq \lambda \leq l$, resulting in both $\hat{z}$ and $x_g$ having a similar set of neighbors for all of the layers as well as the final prediction. However, while this assumption makes intuitive sense, it can be excessively strict for generating adversarial examples. The adversary only needs a large fraction of the neighbors of $\hat{z}$ to be of class $y_{adv}$. By extending the gradient-based attack on \knn, we formulate an analogous optimization problem for attacking \dknn as follows:
\begin{align} \label{eq: opt_dknn_1}
    \hat{\delta} = &\argmin_\delta \sum_{i=1}^{m} \sum_{\lambda=1}^{l} ~w_i \cdot \sigma \big(\norm{f_\lambda(x_i) - f_\lambda(z + \delta)}_2 - \eta_\lambda\big) \\
    &\text{such that} \norm{\delta}_p \leq \epsilon
    \text{ and } z + \delta \in [0, 1]^d \nonumber
\end{align}
The $m$ samples are chosen similarly to the attack on \knn. In the interest of space, we omit the formulation for the $\ell_2$ constraint as it is also analogous to Eq. \ref{eq: opt_knn_3}.


\section{Experimental Setup}

We reimplement \dknn from \papernot with the same hyperparameters, including the network architecture and the value of $k = 75$. We evaluate our attacks on the MNIST dataset \cite{mnist} as past research suggests that finding adversarial examples on other tasks is even easier. 60,000 samples are used as the training samples for \knn, \dknn, as well as the neural network part of \dknn. 750 samples (75 from each digit) are held out as the calibration set, leaving 9,250 test samples for evaluating the accuracy and the robustness of the classifiers against the attacks. Similarly to \papernot, for a quick nearest neighbor search on \dknn, we use a locality-sensitive hash (LSH) from the FALCONN Python library, which is based off cross-polytope LSH by Andoni et al. \cite{andoni15}. \knn uses an exact neighbor search without any approximation. The \knn and the \dknn have an accuracy of 95.74\% and 98.83\% on the clean test set, respectively. The neural network alone has an accuracy of 99.24\%.

All of the attacks are evaluated under both $\ell_2$- and $\ell_\infty$-norm constraints, except for the naive attack on \knn and the mean attacks. For simplicity, we only evaluate untargeted attacks. Both the mean and the naive attacks use only five binary search steps. For the other attacks, we use 400 iterations of gradient updates and five steps of binary search on the $\ell_2$-penalty constant. The Adam optimizer is used in the gradient-based attack, and to save computation time, we only check for the termination condition (i.e., whether $\hat{z}$ is misclassified) three times at iterations 320, 360, and 400, instead of at every step.

We made minimal effort to select hyperparameters. We fix the steepness $\alpha$ of the sigmoid at 4, and for \dknn, we arbitrarily choose the initial $m$ samples to be the $k$ training samples with class $y_{adv}$ whose first-layer representation is closest to that of $z$. For the $\ell_2$-norm attacks, $\epsilon$ is simply chosen to be 0 with the constant $c$ being 1. This choice of penalty generally allows the optimization to find adversarial examples most of the time but may result in unnecessarily large perturbations. To set a more strict constraint, one could set $\epsilon$ to a desired threshold and $c$ to a very large number.


\section{Results} \label{sec: results}

\subsection{k-Nearest Neighbors}

\begin{table}[t]
\centering
\caption{Evaluation of all the attacks on \knn.}
\label{tab:knn}
\begin{tabular}{@{}lccc@{}}
\toprule
 \multicolumn{1}{c}{Attacks} & Accuracy & Mean Distortion in $\ell_2$ \\ \midrule
 Clean Samples                   & 0.9574 & - \\
 Mean Attack                     & \textbf{0.0589} & \textbf{8.611} \\
 Naive Attack                    & 0.7834 & 8.599 \\
 Gradient Attack ($\ell_2$)      & \textbf{0.0989} & \textbf{6.565} \\
 Gradient Attack ($\ell_\infty$) & 0.8514 & 5.282 \\ \bottomrule
\end{tabular}
\end{table}

Table \ref{tab:knn} displays the accuracy and mean $\ell_2$ distortion of the successful adversarial examples for \knn. As expected, the mean attack is very good at finding adversarial examples but the perturbation is large and the adversarial examples sometimes introduce anomalies that may be noticeable to humans. Surprisingly, the naive attack performs  much more poorly compared to the mean attack, indicating that the heuristic used to choose the set of target samples can significantly affect the attack success rate. The gradient-based attack with the $\ell_2$-norm performs well and is on par with the mean attack while having considerably smaller mean distortion. On the other hand, the gradient attack with $\ell_\infty$-norm of 0.2 is mostly unsuccessful. We speculate this might be because $\epsilon=0.2$ is too small and the $\ell_\infty$-norm is an ineffective choice of norm as \knn relies on Euclidean distance in the pixel space for prediction.

\subsection{Deep k-Nearest Neighbors}

\begin{table}[t]
\centering
\caption{Evaluation of all the attacks on \dknn.}
\label{tab:dknn}
\begin{tabular}{@{}lccc@{}}
\toprule
 \multicolumn{1}{c}{Attacks} & Accuracy & Mean Dist. & Mean Cred. \\ \midrule
 Clean Samples                    & 0.9883 & - & 0.6642 \\
 Mean Attack                      & 0.1313 & 4.408 & 0.0172 \\
 \midrule
 Baseline Attack ($\ell_2$)       & 0.1602 & 3.459 & 0.0185 \\
 Baseline Attack ($\ell_\infty = 0.2$)  & 0.8891 & 2.660 & 0.0807 \\
 Baseline Attack (fixed $\ell_2$) & 0.5004 & 3.435 & 0.1385 \\
 \midrule
 Gradient Attack ($\ell_2$)       & \textbf{0.0000} & \textbf{2.164} & 0.0482 \\
 Gradient Attack ($\ell_\infty = 0.2$)  & 0.1744 & 3.476 & 0.1037 \\ 
 Gradient Attack (fixed $\ell_2$) & 0.0059 & 3.375 & \textbf{0.3758} \\ \bottomrule
\end{tabular}
\end{table}

Table \ref{tab:dknn} compares the accuracy, mean $\ell_2$ distortion, and mean credibility of the successful adversarial examples for \dknn between the three attacks. Our novel gradient-based attack outperforms the baseline as well as the mean attack by a significant margin. With an $\ell_\infty$-norm constraint of 0.2, the gradient attack reduces the classifier's accuracy much further compared to the baseline. With an $\ell_2$-norm constraint, our gradient attack also performs better with smaller perturbation. Although the mean attack reduces the accuracy even lower than the gradient attack with $\ell_\infty$-norm of 0.2, it has lower mean credibility and the perturbation is also considerably larger and more visible to humans.

Unlike an $\ell_\infty$ constraint, which is strictly enforced by the change of variables trick, an $\ell_2$ constraint is written as a penalty term with only a tunable weighting constant. To compare the baseline and the gradient attacks under a similar $\ell_2$-norm, we arbitrarily set $\epsilon$ to be the mean $\ell_2$-norm of the $\ell_\infty$ gradient attack (3.476) and the constant $c$ to be just high enough that the optimization still finds successful attacks with a minimal violation on the constraint $\norm{\epsilon}_2 \leq 3.476$. We report the results for both attacks in Table \ref{tab:dknn} on the ``fixed $\ell_2$'' rows. The gradient attack, when given a large $\ell_2$ budget, can increase the credibility significantly and reduce the accuracy to almost zero (0.6\%). In contrast, the baseline attack can only find adversarial examples for about 50\% of the samples under the same $\ell_2$ constraint. 

\begin{figure*}[t]
  \centering
  \includegraphics[width=\textwidth]{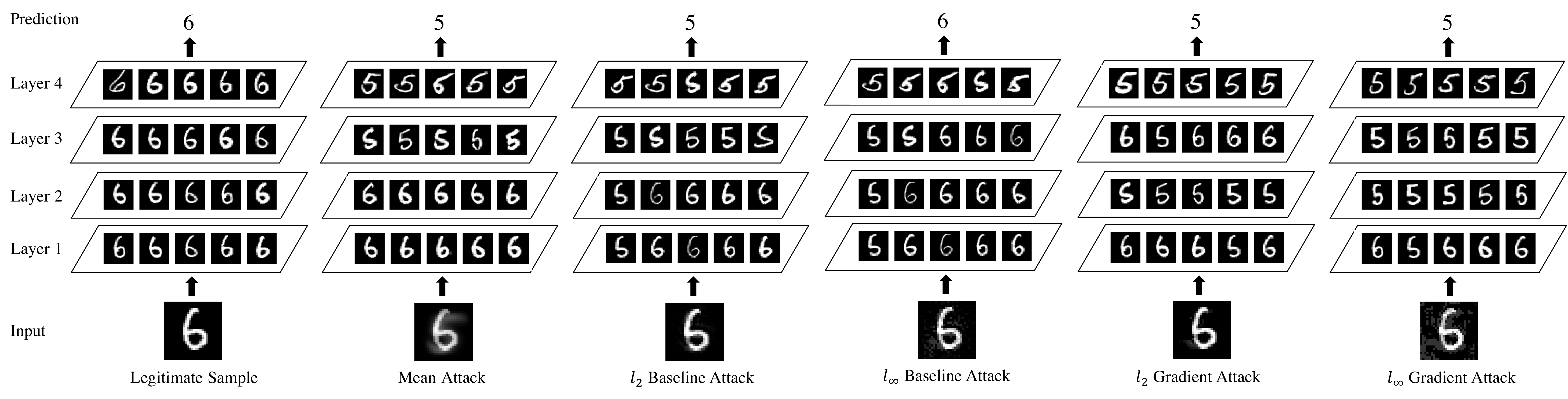}
  \caption{Each column shows five nearest neighbors for each of the four deep representation spaces of \dknn. From left to right, the inputs are a randomly chosen legitimate sample, its $\ell_2$ and $\ell_\infty$ baseline attacks, and its $\ell_2$ and $\ell_\infty$ gradient attacks. For the $\ell_\infty$-norm constraint, $\epsilon$ is 0.2. The legitimate sample is correctly predicted by the \dknn, and all of the attacks succeed in changing the prediction from a six to a five, except for the $\ell_\infty$ baseline attack.}
  \label{fig:grad_attack}
  \vspace{-5pt}
\end{figure*}

Fig. \ref{fig:grad_attack} shows a clean sample and its adversarial versions generated by all of the attacks along with their five nearest neighbors at each of the four layers of representation. On the first column, all of the 20 neighbors of the clean sample have the correct class (a six). On the other hand, the majority of neighbors of the adversarial examples are of the incorrect class (a five) with an exception of the first layer whose neighbors generally still come from the correct class. The other common property of all the attacks is that almost every neighbor in the final layer has the adversarial class.

Note that the $\ell_2$-attacks, both the baseline and the gradient-based attack, often perturb the sample in a semantically meaningful manner. Most are subtle, but some are quite prominent. For instance, the input of the third column from the left in Fig. \ref{fig:grad_attack} is perturbed by slightly removing the connected line that distinguishes between a five and a six, making the adversarial example appear somewhat ambiguous to humans. In contrast, the $\ell_\infty$ adversarial examples usually spread the perturbation over the entire image without changing its semantic meaning in a way that is noticeable to humans.

\begin{figure}[t]
    \centering
    \begin{subfigure}[t]{0.24\textwidth}
        \centering
        \includegraphics[width=\textwidth]{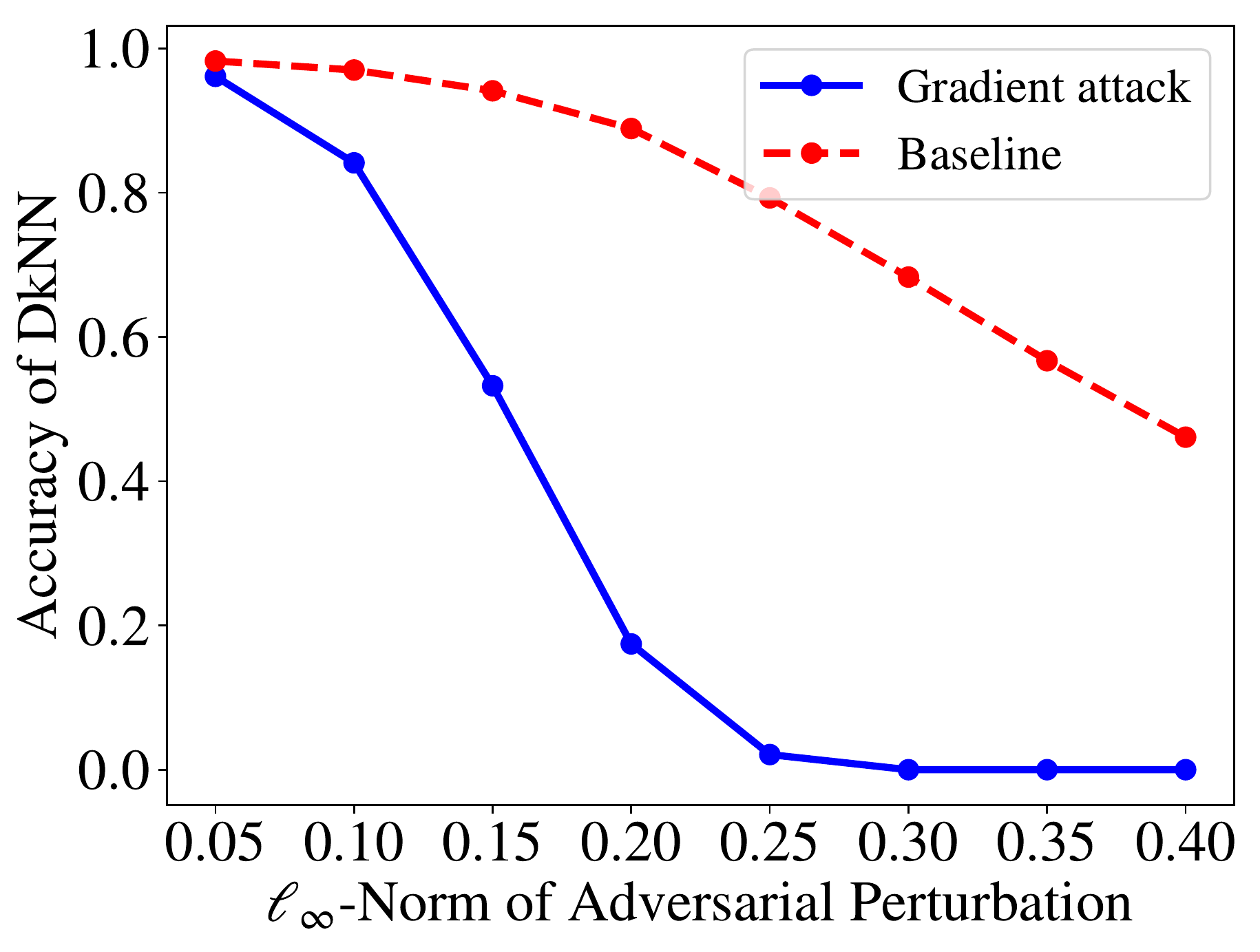}
    \end{subfigure}%
    ~ 
    \begin{subfigure}[t]{0.24\textwidth}
        \centering
        \includegraphics[width=\textwidth]{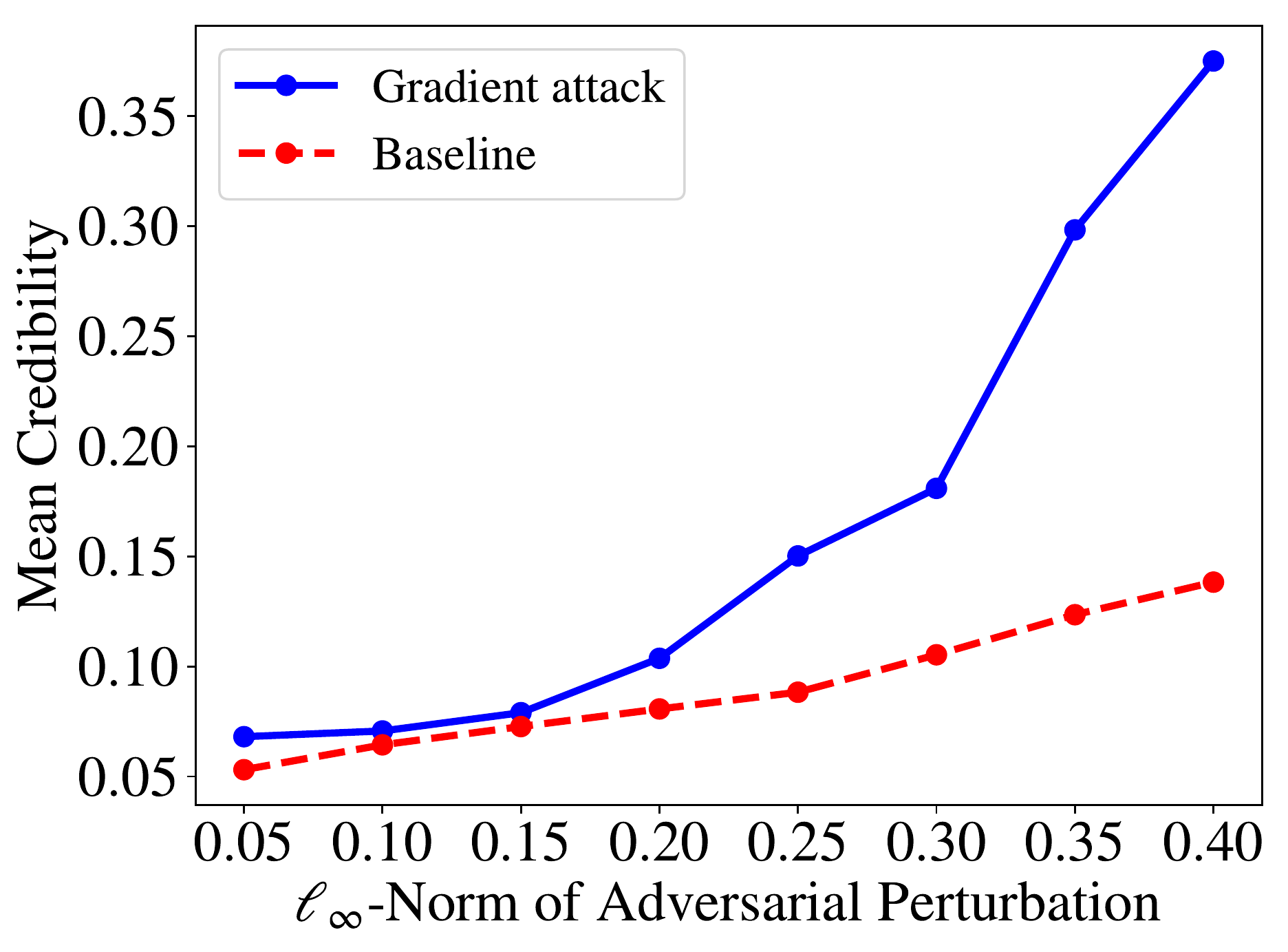}
    \end{subfigure}
    \caption{(a) Accuracy and (b) mean credibility of \dknn under the baseline attack and our gradient-based attack at different $\ell_\infty$-norm constraints.}
    \label{fig:grad_baseline}
    \vspace{-5pt}
\end{figure}


For the $\ell_\infty$-norm constraint, as we increase $\epsilon$, the accuracy of \dknn drops further and hits zero at $\epsilon = 0.3$, as shown in Fig. \ref{fig:grad_baseline}(a), whereas increasing $\epsilon$ on the baseline attack reduces accuracy at a much slower rate.


Fig. \ref{fig:grad_baseline}(b) displays the mean credibility of successful adversarial examples generated from the baseline and the gradient attacks. As expected, as we increase $\epsilon$, the mean credibility also increases for both attacks because the adversarial example can move closer to training samples from the target class. The gradient-based attack increases the mean credibility at a much faster rate than the baseline potentially because its objective function indirectly corresponds to the credibility as it takes into account $m$ training samples instead of one like the baseline. In the next section, we discuss the possibility of detecting adversarial examples by setting a threshold on the credibility score.


\section{Discussion}

\subsection{Credibility Threshold} \label{ssec:cred_thres}

\begin{figure}[t]
  \centering
  \includegraphics[width=0.4\textwidth]{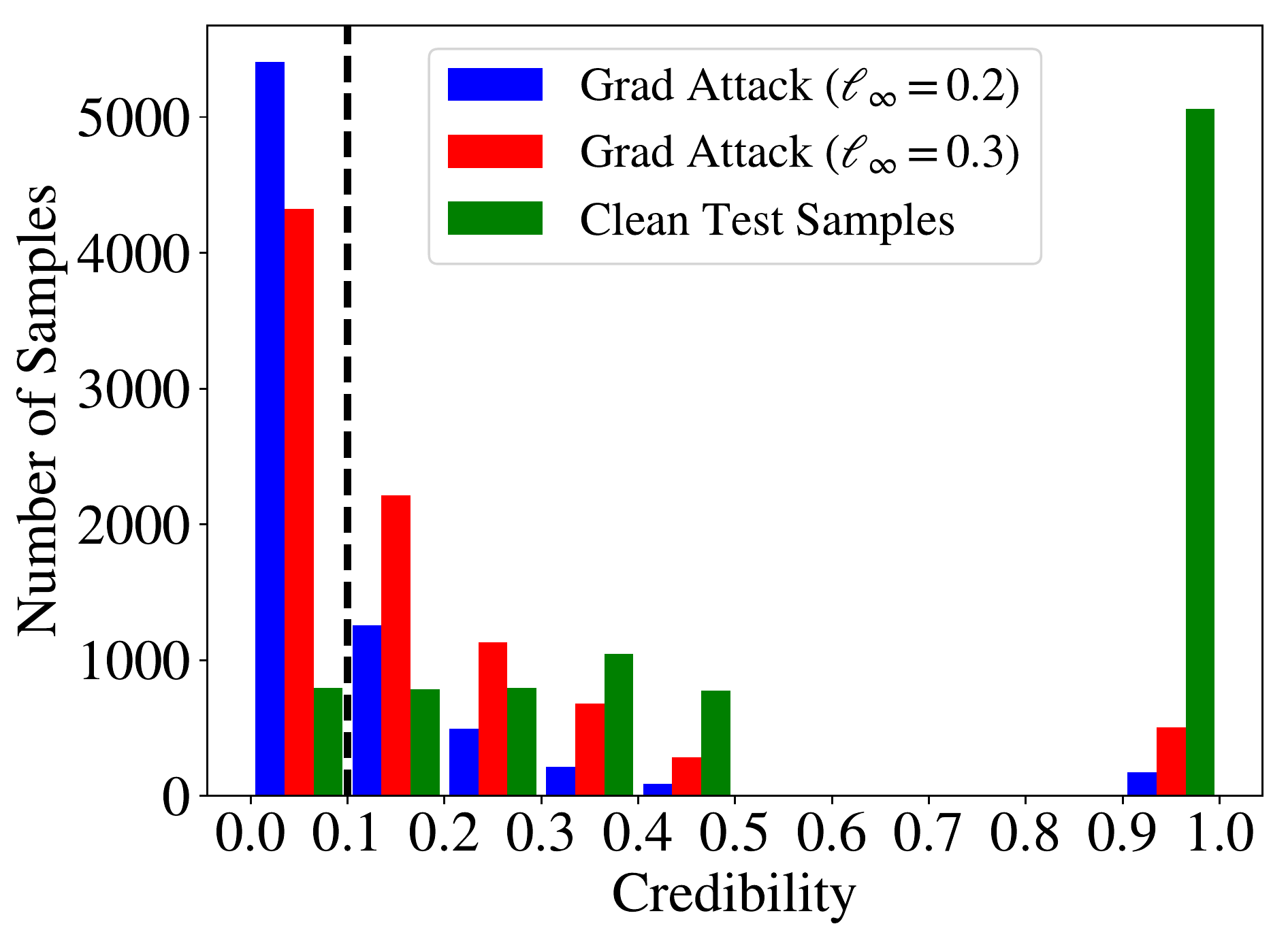}
  \caption{Histogram of credibility of the clean test samples and the adversarial examples generated from the gradient-based attack with the $\ell_\infty$-norm constraint of 0.2 and 0.3. The black dashed vertical line indicates credibility of 0.1.}
  \label{fig:hist}
  \vspace{-5pt}
\end{figure}

\papernot argues that the credibility output by \dknn is a well-calibrated metric for detecting adversarial examples. In Fig. \ref{fig:hist}, we show the distribution of the credibility for the clean test set and for adversarial examples generated from the gradient-based attack with two different $\ell_\infty$-norms. Most of the test samples (around 55\%) have credibility between 0.9 and 1. On the other hand, the majority of the adversarial examples have credibility less than 0.1, suggesting that setting a threshold on credibility can potentially filter out most of the adversarial examples. However, doing so comes at a cost of lowering accuracy on legitimate samples. Choosing a credibility threshold of 0.1 reduces accuracy on the test set to 91.15\%, which is already very low for MNIST, and with this threshold, 28\% and 43\% of the adversarial examples with $\ell_\infty$-norm of 0.2 and 0.3 respectively still pass the threshold and would not be detected. It is also important to note that our attack is not designed to maximize the credibility. Rather, it is designed to find adversarial examples with minimal distortion. Simple parameter fine-tuning, e.g. a larger $m$, more iterations, and a smaller $\eta$, might all help increase the credibility.

Our experiments suggest that \dknn's credibility may not be sufficient for eliminating adversarial examples, but it is still a more robust metric for detecting adversarial examples than a softmax score of typical neural networks. Unfortunately, thresholding the credibility hurts accuracy on legitimate examples significantly even for a simple task like MNIST. According to \papernot, the SVHN and GTSRB datasets both have a larger fraction of legitimate samples with low credibility than MNIST, making a credibility threshold even less attractive. Experiments with the ImageNet dataset, deeper networks, choosing which layers to use, and pruning \dknn for robustness are all interesting directions for future works.




\section{Conclusion}

We propose two heuristic attacks and a gradient-based attack on \knn and use them to attack \dknn. We found that our gradient attack performs better than the baseline: it generates adversarial examples with a higher success rate but lower distortion on both $\ell_2$ and $\ell_\infty$ norms. Our work suggests that \dknn is vulnerable to adversarial examples in a white-box adversarial setting. Nonetheless, \dknn still holds promise as a direction for providing significant robustness against adversarial attacks as well as interpretability of deep neural networks.

\section{Acknowledgements}

This work was supported by the Hewlett Foundation through the Center for Long-Term Cybersecurity and by generous gifts from Huawei and Google.

\bibliographystyle{IEEEtran}
\bibliography{advex,advml,ml,interpret}

\end{document}